\documentclass[twocolumn,showpacs,preprintnumbers,amsmath,amssymb]{revtex4}
\usepackage{epsfig}

\usepackage{graphicx}% Include figure files
\usepackage{dcolumn}% Align table columns on decimal point
\usepackage{bm}% bold math

\newcommand{\be}{\begin{equation}}
\newcommand{\ee}{\end{equation}}

\newcommand{\bea}{\begin{eqnarray}}
\newcommand{\eea}{\end{eqnarray}}
\newcommand{\beq}{\begin{equation}}
\newcommand{\eeq}{\end{equation}}
\newcommand{\nn}{\nonumber}

\def\ga{\mathrel{\mathpalette\fun >}}
\def\fun#1#2{\lower3.6pt\vbox{\baselineskip0pt\lineskip.9pt
\ialign{$\mathsurround=0pt#1\hfil##\hfil$\crcr#2\crcr\sim\crcr}}}

\begin{document}

\title{Hadron diffractive scattering
at ultrahigh energies, real part of the amplitude
and Coulomb interaction}

\author{V.V. Anisovich}
\affiliation{National Research Centre "Kurchatov Institute",
Petersburg Nuclear Physics Institute, Gatchina 188300, Russia}
\author{V.A. Nikonov}
\affiliation{National Research Centre "Kurchatov Institute",
Petersburg Nuclear Physics Institute, Gatchina 188300, Russia}
\author{J. Nyiri}
\affiliation{Institute for Particle and Nuclear Physics, Wigner RCP,
Budapest 1121, Hungary}

\date{\today}

\begin{abstract}
On the basis of requirements of unitarity and analyticity
we analyze the real and imaginary parts of the $pp^\pm$ scattering
amplitudes at recent ultrahigh energies, 1-100 TeV. The predictions
for the region  $\sqrt s\, >\, 100$ TeV and ${\bf q}^2<0.4$ GeV$^2$ are given
supposing the black disk asymptotic regime. It turns out that the real
part of the amplitude is concentrated in the impact parameter space at the
border of the black disk. The interplay of hadron and Coulomb interactions
is discussed in terms of the $K$-matrix function.
The $pp$ diffractive scattering cross section at 7 TeV is calculated with
Coulomb interaction taken into account.

  \end{abstract}
\pacs{ 13.85.Lg, 13.85.-t, 13.75.Cs, 14.20.Dh}
\maketitle

\section{Introduction}
Ultrahigh energy proton-proton interaction data at LHC
\cite{totem,atlas,cms,alice} and the cosmic ray data \cite{auger}
definitely tell us that at energies $\sqrt s\sim 5-50$ TeV the asymptotic
regime is switching on for diffractive scattering processes.
These data together with those of ISR \cite{pre} demonstrate the steady growth
of $\sigma_{tot}$, $\sigma_{el}$ and $\sigma_{inel}$ with the energy
increase, a shrinkage of the diffractive cone in $d\sigma_{el}/d{\bf q}^2$
and a relative supression of the real part of the scattering amplitude
%, $A_{\Re}/A_{\Im}<<1 $ at ${\bf q}^2\simeq 0$.

The growth of the cross sections $\sigma_{tot}$, $\sigma_{el}$, $\sigma_{inel}$ and the
diffractive scattering slope is consistent with the picture of fast moving
hadrons as parton clouds with an increasing transverse size. The
gluonic origin of the parton cloud explicates a slow growth of the
diffractive scattering slope and a late start of the asymptotic regime:
the effective mass of the soft gluon is notsmall being of the order of
$800-1000$ MeV \cite{parisi,field}.

The observed growth of total cross sections at preLHC energies \cite{serp,fermi}
initiated studies of models with the
supercritical pomeron \cite{cape,volk,kaid,donn}.
The discussion of the power growth of cross sections with energy
actualized  the problem of $s$-channel unitarization of scattering
amplitudes and the use the Glauber approach \cite{glau}. Taking into
account the $s$-channel rescatterings, the power-$s$ growth of amplitudes
is dampened to the $(\ln^2 s)$-type \cite{Gaisser,Block,Fletcher}, to the
limits of the Froissart bound \cite{Froi}. Still, let us emphasize that
exceeding it does not violate the general constraints for the scattering
amplitude \cite{azimov}.

The black disk picture appears to be a rather natural mode for the ultrahigh
energy corresponding to non-coherent parton interactions in hadron collisions.
For the black disk scenario the profile function at $\sqrt{s}\ga 100$ TeV
gets frozen inside the disk area, $T(b)\simeq 1$ at $b< R_{black\;disk}$,
while the increasing radius of the black disk, $R_{black\;disk}$,
determines the total, elastic and inelastic cross sections:
$\sigma_{tot}\simeq 2\pi R^2_{black\;disk}$,
$\sigma_{el}\simeq \pi R^2_{black\;disk}$
and $\sigma_{inel}\simeq \pi R^2_{black\;disk}$. The black disk mode was
intensely discussed in the last decade, see, for example,
\cite{DN,1110.1479,sch-rys,1202.2016,1208.4086,ann1,ann2} and
references therein.

In the present paper we underline that at asymptotic energies the
imaginary part of the amplitude $ A_{\Im}({\bf q}^2,\ln s)$ turns into a
generating function for the real part, $A_{\Re}({\bf q}^2,\ln s)$, due to
unitarity and analyticity requirements. In Section 2
the real parts of the hadronic scattering amplitude are calculated for a set
of energies, $\sqrt{s}=1,10,10^2,...,10^6$ TeV, and profile functions,
$T_\Im(b)$ and $T_\Re(b)$, are presented. In Section 3 we discuss
a combined action of the Coulomb and hadronic interactions for the
diffractive scattering region. If here the eikonal approach works, the
straightforward way to take into account the interplay of these interactions
is the use of the $K$-matrix function technique, thus keeping valid the
unitarity condition. We present the corresponding formulae
and calculate the $pp$ diffractive scattering cross section at 7 TeV with
the Coulomb interaction taken into account.

\section{Real part of the hadronic scattering amplitude }

The high energy scattering ampitude is dominantly imaginary, the real part
of the amplitude is a next-to-leading term. So, in the pre-asymptotic
region we should include into consideration of the scattering
amplitude  both the imaginary and real parts.
The amplitude reads:
\bea \label{c1}
A({\bf q}^2 ,\xi)&=& \int d^2b e^{i{\bf q}{\bf b}}T(b,\xi),
\label{ampl}
\nn \\
T(b,\xi)&=&T_{\Im}(b,\xi)-iT_{\Re}(b,\xi)\,,   \eea
where $\xi=\ln s$, $b=|{\bf b}|$. For the profile function we write:
\be
T(b,\xi)=1-\eta(b,\xi)\exp{(2i\delta(b,\xi)})
= \frac{-2i K(b,\xi)}{1-iK(b,\xi)}\,,
\label{tkmatr}
\ee
presenting it in terms of the phase shift $\delta(b,\xi)$ and inelasticity
parameter $\eta(b,\xi)$, or the $K$-matrix function $K(b,\xi)$.

\begin{figure*}[ht]
%\Fig. 1
\centerline{\epsfig{file=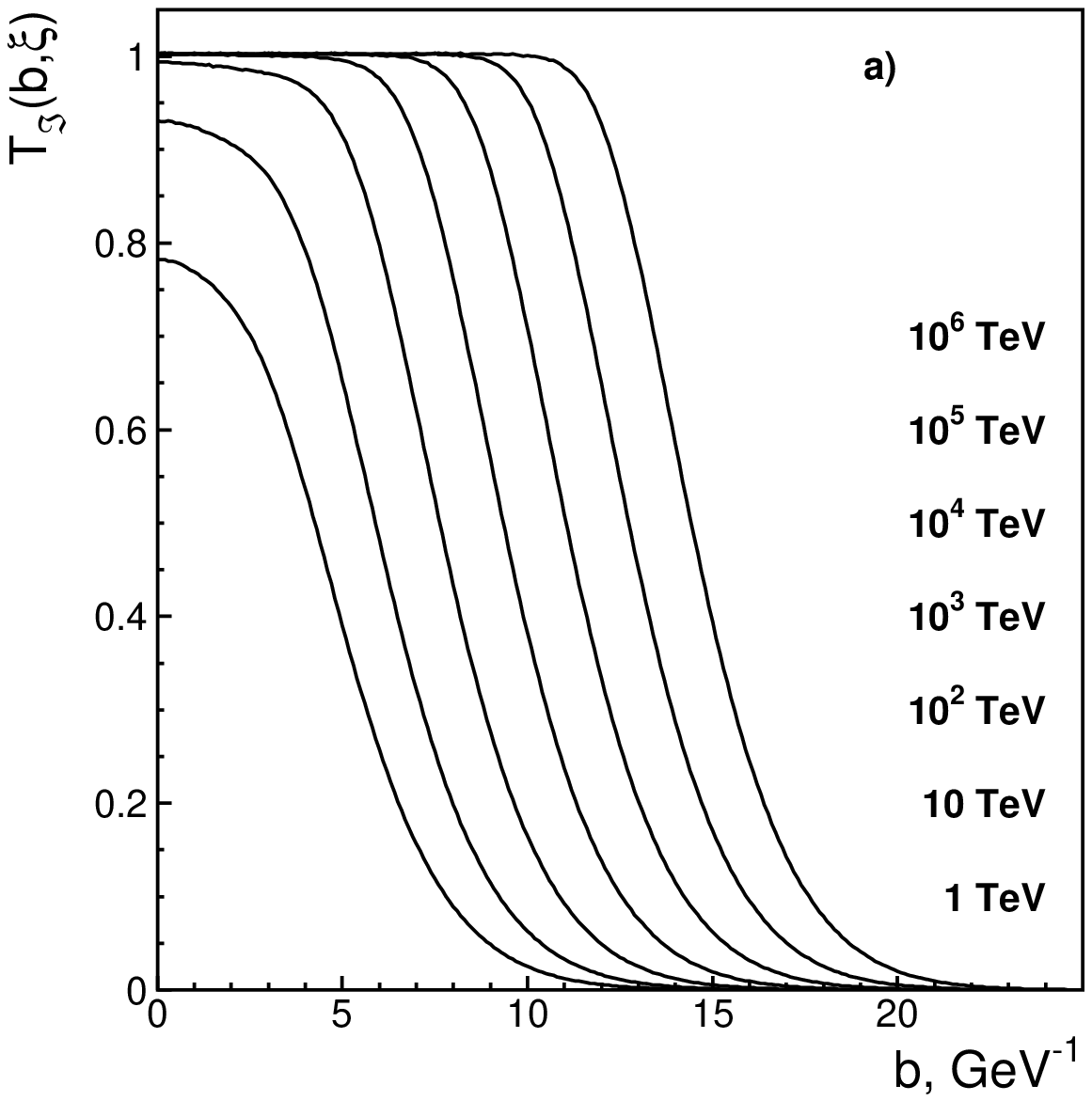,width=6cm}
            \epsfig{file=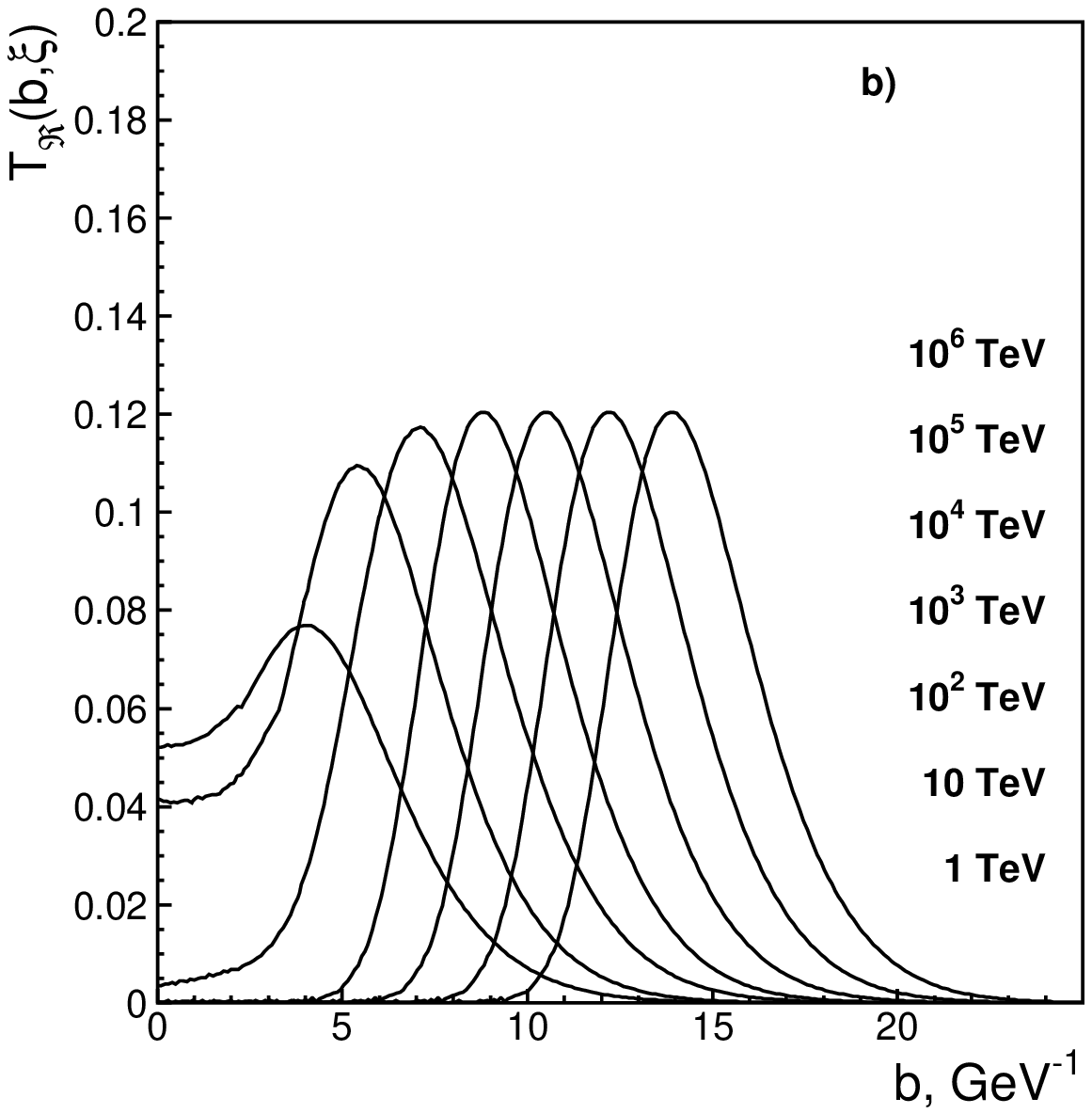,width=6cm}}
\caption{
    Profile functions for imaginary and real parts of the $pp$ scattering amplitude,
    $T_{\Im}(b,\xi)$ and $T_{\Re}(b,\xi)$,
  at a set of energies $\sqrt{s}=1,10,10^2,...,10^6$ TeV; for
    $\sqrt{s}>100$ TeV the black disk mode is suggested.
%; Fig. 1a  is taken from ref. \cite{ann2}.
}
\label{f1}
\end{figure*}

\begin{figure*}[ht]
%\Fig. 2
\centerline{\epsfig{file=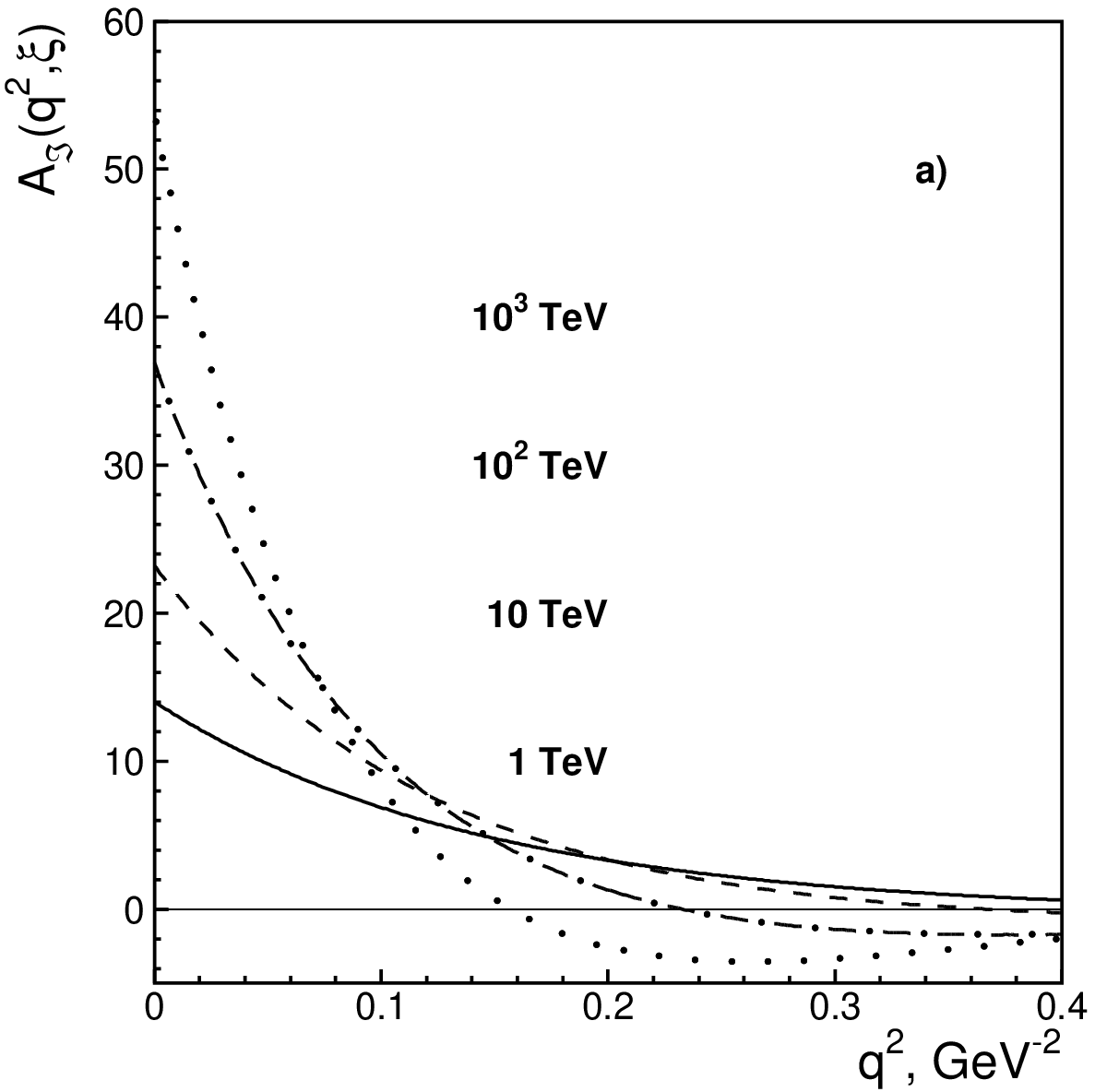,width=6cm}
            \epsfig{file=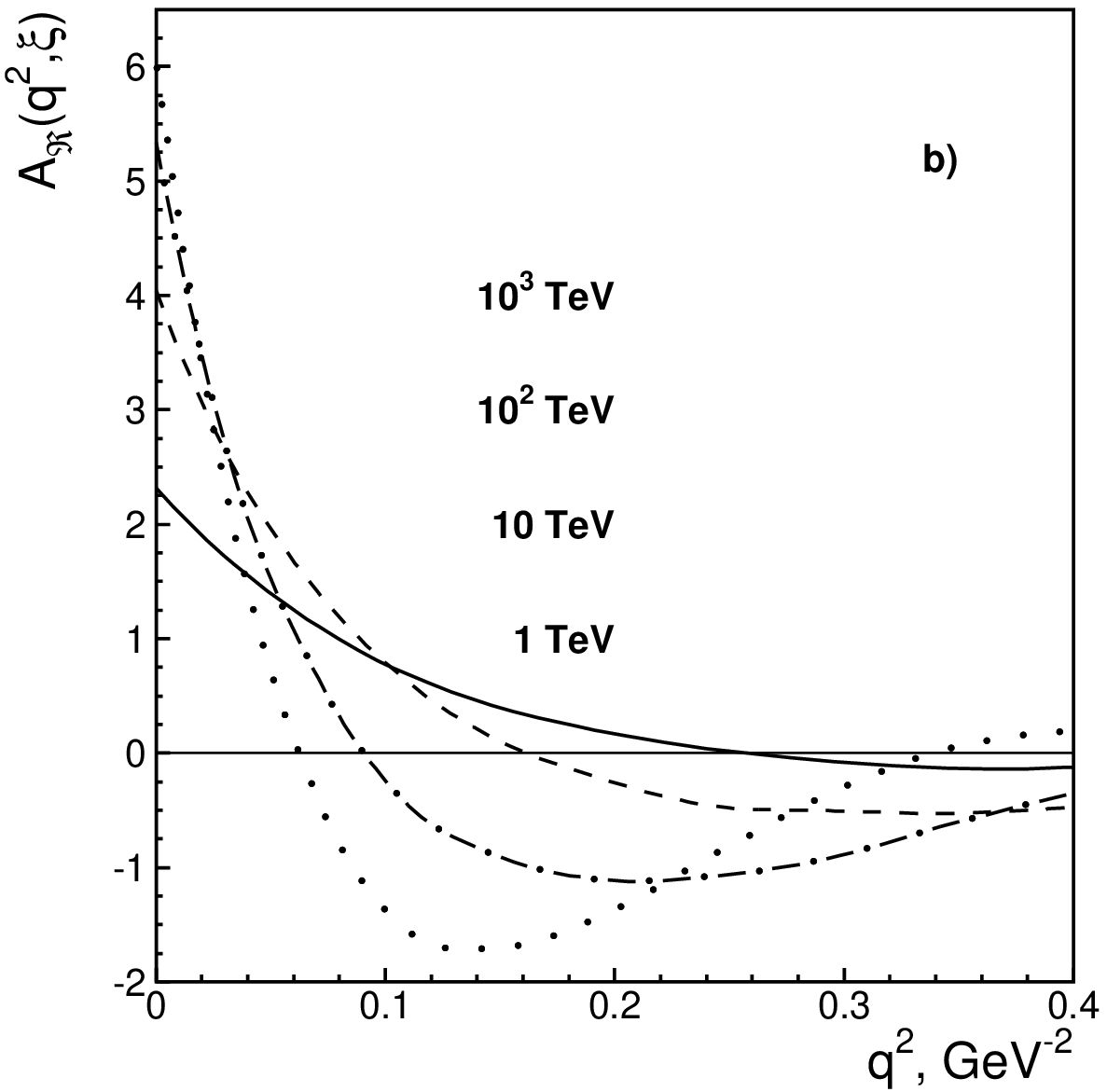,width=6cm}}
\caption{
     Imaginary and real parts of the $pp$ scattering amplitude,
     $A_{\Im}(q^2,\xi)$ and $A_{\Re}(q^2,\xi)$,
  at  energies $\sqrt{s}$=1 (solid),10 (dashed), 10$^2$ (dot-dashed),
  10$^3$  (dotted) TeV;
for  $\sqrt{s}>>10$ TeV the black disk mode is suggested.}
\label{f2}
\end{figure*}

If the imaginary part of the amplitude dominates, the next-to-leading
terms include real part due to analyticity requirement. Namely,
we should take into account contributions both of the $s$-channel and
the $u$-channel. Suggesting
$\sigma_{tot}(pp)=\sigma_{tot}(p\bar p)$ at $s\to \infty$ we write:
\bea  \label{c2}
&&\frac12\Big[A({\bf q}^2,\ln s)+A({\bf q}^2,\ln(- s))\Big]\simeq
\nn \\
&&\simeq A({\bf q}^2,\ln s)+
\frac{\partial A({\bf q}^2,\ln s)}{\partial(\ln s)}
\cdot\frac{-i\pi}{2}
\nn \\
&&\simeq
A_{\Im}({\bf q}^2,\ln s)+
\frac{\partial A_{\Im}({\bf q}^2,\ln s)}{\partial (\ln s)}
\cdot\frac{-i\pi}{2}\,,
\eea
where $\ln(-s)=\ln s -i\pi$. It means the amplitude $A_{\Im}({\bf q}^2,\ln s)$
is the generating function for $A_{\Re}({\bf q}^2,\ln s)$. Analogously we
write for the profile function:
\be  \label{c3}
  T_{\Re}(b,\ln s)
\simeq\frac{\pi}{2}\frac{\partial T_{\Im}(b,\ln s)}{\partial (\ln s)} \,.
\ee
The usual
notation reads $T_{\Re}(b,\xi)/T_{\Im}(b,\xi)=\rho(b,\xi)$, therefore
the total and elastic cross sections
are written as:
\bea
T(b,\xi)&=&(1+i\rho)T_{\Im}(b,\xi)\,,
\nn \\
\sigma_{tot}&=&2 \int d^2b T_{\Im}(b,\xi),  \nn \\
4\pi\frac{d\sigma_{el}}{d{\bf q}^2}&=&
(1+\rho^2)A_{\Im}^2({\bf q}^2)\,.
\eea
Taking into account that $\rho^2$ is small,  $\rho^2\sim 0.01$, one can
approximate:
\be \label{c5}
\Big|A_{\Im}({\bf q}^2 ,\xi)\Big| \simeq
2\pi^\frac12\sqrt{
\frac{d\sigma_{el}}{d{\bf q}^2}
},
\ee
that make possible direct calculations of the real part of the scattering
amplitude, $A_{\Re}({\bf q}^2 ,\xi)$, on the basis of the energy dependence
of the diffractive scattering cross section.

\begin{figure*}[ht]
%\Fig. 3
\centerline{\epsfig{file=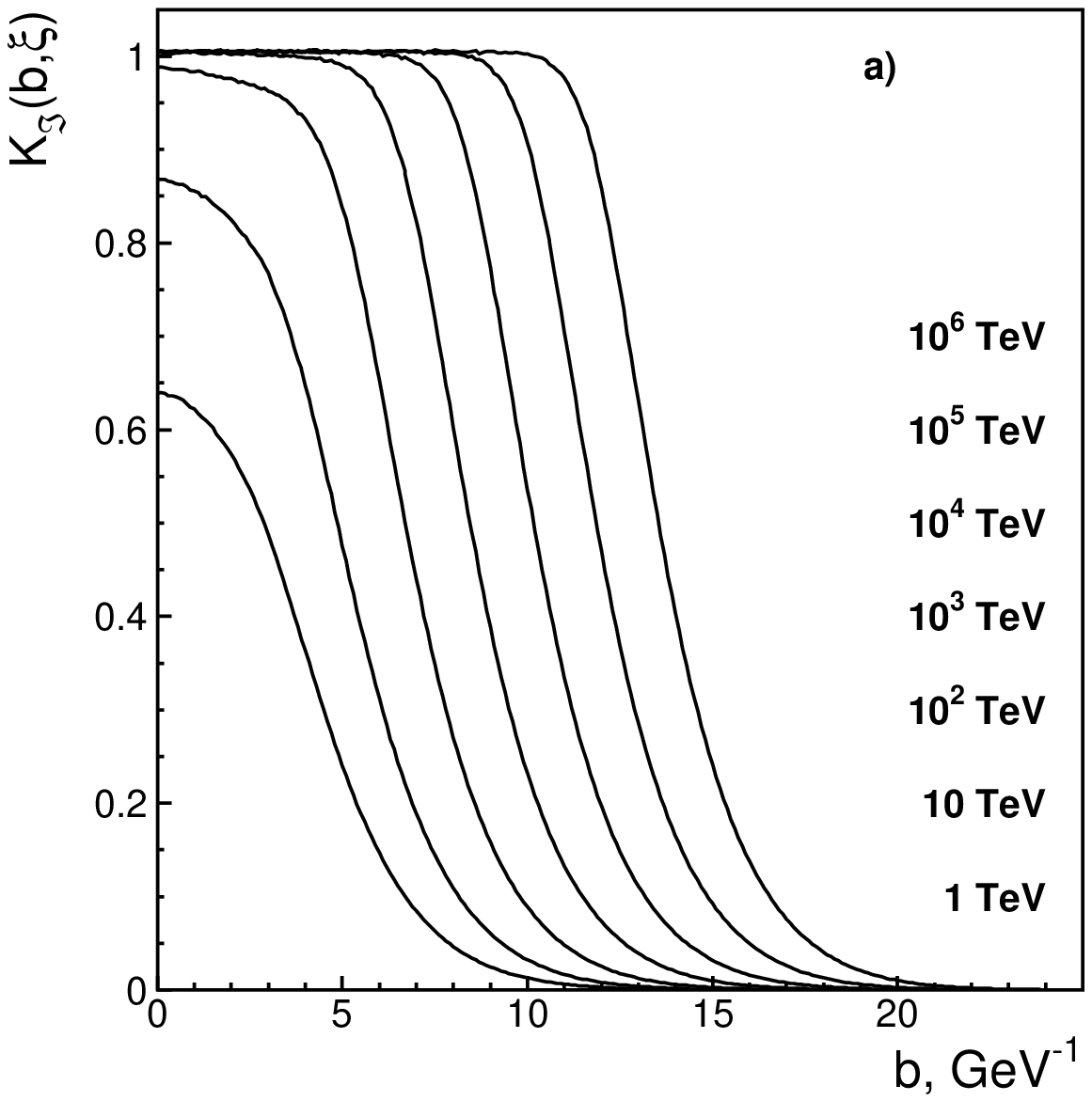,width=6cm}
            \epsfig{file=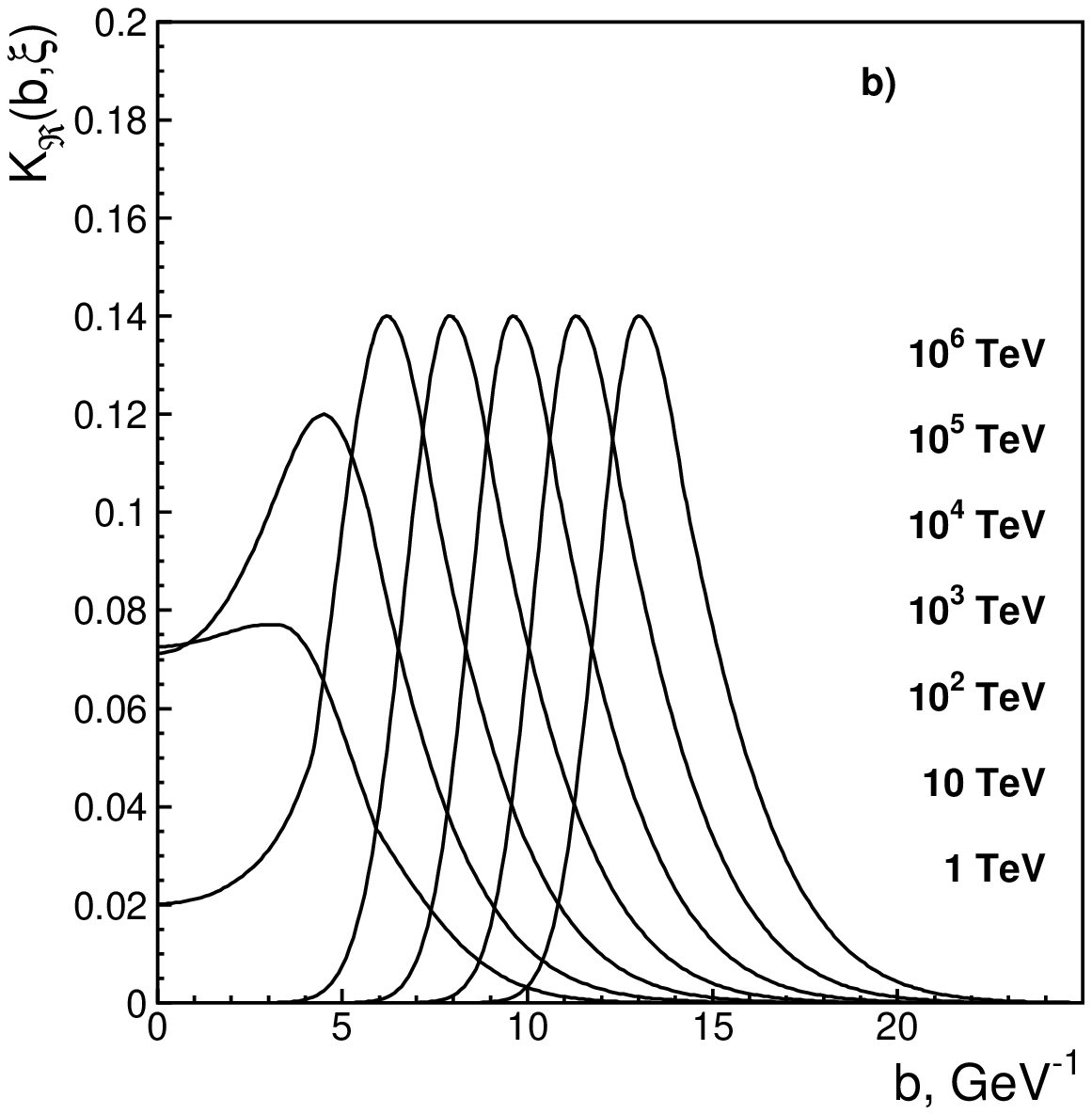,width=6cm}}
\caption{
     Imaginary and real parts of the $K$-matrix function,
     $K_{\Im}(b,\xi)$ and $K_{\Re}(b,\xi)$,
  at  energies $\sqrt{s}=1,10,10^2,...,10^6$ TeV.
%;  Fig. 1a is taken from ref. \cite{ann2}.
}
\label{f3}
\end{figure*}

\subsection{Calculation of the real part of the scattering
amplitude}

Using Eq.~(\ref{c5}) we calculate $A_{\Re}({\bf q}^2 ,\xi)$  and,
correspondingly, the profile functions $T_{\Re}(b,\xi)$ and $T_{\Im}(b,\xi)$,
see  Figs. \ref{f1}, \ref{f2}.
The figure \ref{f1}a for $T_{\Im}(b,\xi)$ is taken from \cite{ann2} where
a comparison with data is given as well.
Calculations at  $\sqrt{s}\ga 100$ TeV are done suggesting the black disk mode
as a realization of the asymptotic regime.

An advantage of the $K$-matrix function technique is the separation of the
two-particle rescattering states which turn out to be mass-on-shell for
leading terms of the amplitude \cite{amnn}. Following Eq.~(\ref{tkmatr}),
we write:
\be
-iK(b ,\xi)=\frac{T(b ,\xi)}{2-T(b ,\xi)}
\equiv
K_\Im(b ,\xi)-iK_\Re (b ,\xi)\ .
\ee
The functions $K_\Re(b ,\xi)$ and $K_\Im(b ,\xi)$ for
$\sqrt{s}=1,10,10^2,..., 10^6$ TeV are shown in Fig. \ref{f3}.

\subsection{Eikonal approach for scattering amplitude at ultrahigh energies
and the Feynman diagram technique}

For the $pp$-scattering amplitude
$A_{pp\to pp}(pp_{in}\to pp_{out})$ the reproducing
integral reads:
\bea \label{cp-3}
&&
A_{pp\to pp}\left(pp_{in}\to pp_{out}\right)=
K_{pp\to pp}\left(pp_{in}\to pp_{out}\right)
\nn \\
&&
+\int\frac{d^4k_{2'}}{(2\pi)^4i}\;
A_{2\to 2}\left(pp_{in}\to 1'2'\right)
\nn \\
&&
\times
\frac{K_{pp\to pp}\left({1'}{2'}\to
pp_{out}\right)}{(m^2-k^2_{1'}-i0)(m^2-k^2_{2'}-i0)},
  \eea
  to be definite we consider proton-proton scattering.
Here $K_{pp\to pp}$ is the block without two-particle states thus
being up to factor the $K$-matrix function, indices $(1',2')$ refer to
protons in the intermediate state.

Let us consider the scattering amplitude in the cm-system
where we write for the initial protons:
$p_1\equiv (p_0,{\bf p}_\perp,p_z)=(p+m^2/2p,0,p)$ and
$p_2=(p+m^2/2p,0,-p)$.
For intermediate and final state protons we have:
\bea
\label{cp-4}
&&
{\bf k}_{1'\perp}+{\bf k}_{2'\perp}=0,\quad
{\bf k}_{1\perp}+{\bf k}_{2\perp}=0.  \nn
\\
&&
q^2_{2'}=(p_2-k_{2'})^2\simeq-{\bf k}^2_{2'\perp},
\nn \\
&&
q^2_{2'2}=(k_{2'}-k_{2})^2\simeq
-({\bf k}_{2\perp}-{\bf k}_{2'\perp})^2\,,
\eea
where $k_1,k_2$ refer to momenta of outgoing protons.

At ultrahigh energies the $K$-matrix function is dominantly imaginary
for the black disk and resonant disk modes \cite{ann2,Anisovich:2014wha}. That
means the dominance of the mass-on-shell contribution in the loop diagrams.
For the rescattering diagrams this is realized in the replacement:
  \bea \label{cp-5}
&&
\Big[(m^2-k^2_{1'}-i0)(m^2-k^2_{2'}-i0)\Big]^{-1}\to
\nn \\
&&-2\pi^2\delta(m^2-k^2_{1'})\delta(m^2-k^2_{2'})= \nn
\\
&&
=-2\pi^2\delta\left(k_{1'}^{(+)}k_{1'}^{(-)}
-(m^2+{\bf k}^2_{1'\perp})\right) \nn \\
&&\times
\delta\left(k_{2'}^{(+)}k_{2'}^{(-)}
-(m^2+{\bf k}^2_{2'\perp})\right)\;,
\eea
where
$k^{(+)}=k_0+k_z,\quad k^{(-)}=k_0-k_z$.
Then the right-hand side of Eq.~(\ref{cp-3}) reads:
\bea \label{cp-6}
&&
A_{2\to 2}({\bf k}^2_{2\perp},\xi)=
K_{2\to 2}\left({\bf k}_{2\perp}^2,\xi\right)+
\\
&&
+\!\int\frac{d^2k_{2'\perp}}{(2\pi)^2}
A_{2\to 2}({\bf k}^2_{2'\perp},\xi) \frac{i}{4s}
K_{2\to 2}\left(({\bf k}_{2'\perp}-{\bf k}_{2\perp})^2,\xi\right),
\nn
\eea
here $K_{2\to 2}({\bf k}^2_{\perp},\xi)/(4s)= K({\bf k}^2_{\perp},\xi)$ is the
$K$-matrix function in momentum representation.

\begin{figure*}[ht]
%\Fig. 4
\centerline{\epsfig{file=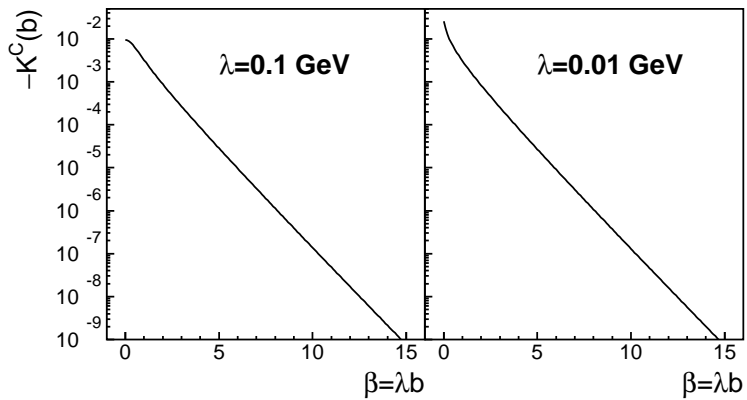,width=10cm}}
\caption{
  Pure coulombic interaction in $pp$ collision at different $\lambda$: the function
  $K^C(b)$ in hadronic region
  of the impact parameters, $b< 25$ GeV$^{-1}$, and
  coulombic region, $\beta= \lambda b>1$.
}
\label{f4}
\end{figure*}

\subsection{Impact parameter presentation}

The Fourier transform
gives the $K$-matrix function in the
impact parameter space:
\bea
\label{cp-7}
&&
\frac{1}{4s}
K_{2\to 2}\left({\bf k}^2_{\perp},\xi\right)=
\int d^2b\exp(i{\bf k}{\bf b})K(b,\xi)\,,
\nn \\
&&
\frac{i}{4s}
A_{2\to 2}\left({\bf k}^2_{\perp},\xi\right)=
\int d^2b\exp(i{\bf k}{\bf b})a(b,\xi)\,,
\eea
Equation (\ref{cp-6}) in the impact parameter space is written as:
\be \label{cp-8}
  a(b,\xi)=
iK(b,\xi)+a(b,\xi)\;
iK(b,\xi)\,.
\ee
Thus, we have the formula of the eikonal approach:
\be \label{cp-9}
a(b,\xi)=
\frac{iK(b,\xi)}{1-iK(b,\xi)}.
\ee
  The function $K(b,\xi)$ depends on the energy and realizes effectively
the interaction which manifests itself in the shrinking of diffractive
cones with the energy increase.

\begin{figure*}[ht]
%\Fig. 5
\centerline{\epsfig{file=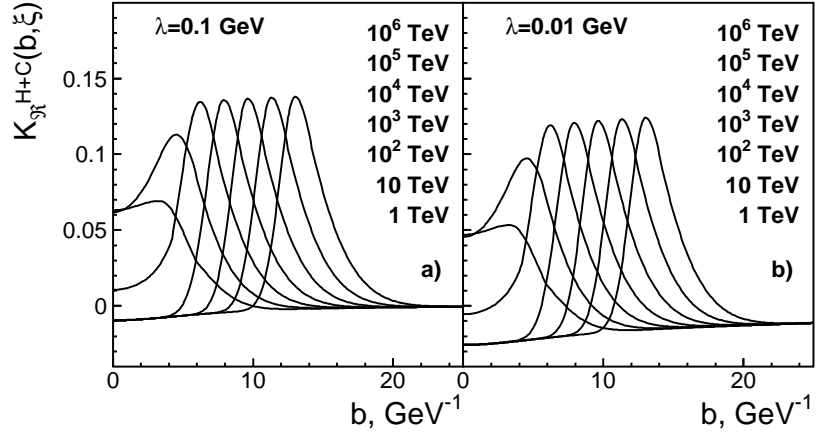,width=11cm}
}
\caption{
  Combined hadronic and coulombic interaction for the $pp$ scattering:
  Real parts of the  $K$-matrix functions at different $\lambda$
  at a set of energies $\sqrt{s}=1,10,10^2,...,10^6$ TeV; for
    $\sqrt{s}>100$ TeV the black disk mode is suggested.
}
\label{f5}
\end{figure*}

\section {Diffractive scattering amplitude
at ultrahigh energies
and Coulomb interaction }

The interplay of hadronic and Coulomb interactions was studied in a set of
papers, see \cite{bethe,solo,west,fran,coha,kund} and references therin.
At ultrahigh energies and small ${\bf q}^2$,  where the eikonal works,
the straightforward way to take into account the combined action of hadronic
and Coulomb interactions ($H+C$) is to use the technique of the $K$-matrix
function. Here we illustrate this way by corresponding calculations of
$K^{H+C}(b,\xi)$ and profile function $T^{H+C}(b,\xi)$.

\begin{figure*}[ht]
%\Fig. 6
\centerline{
  \epsfig{file=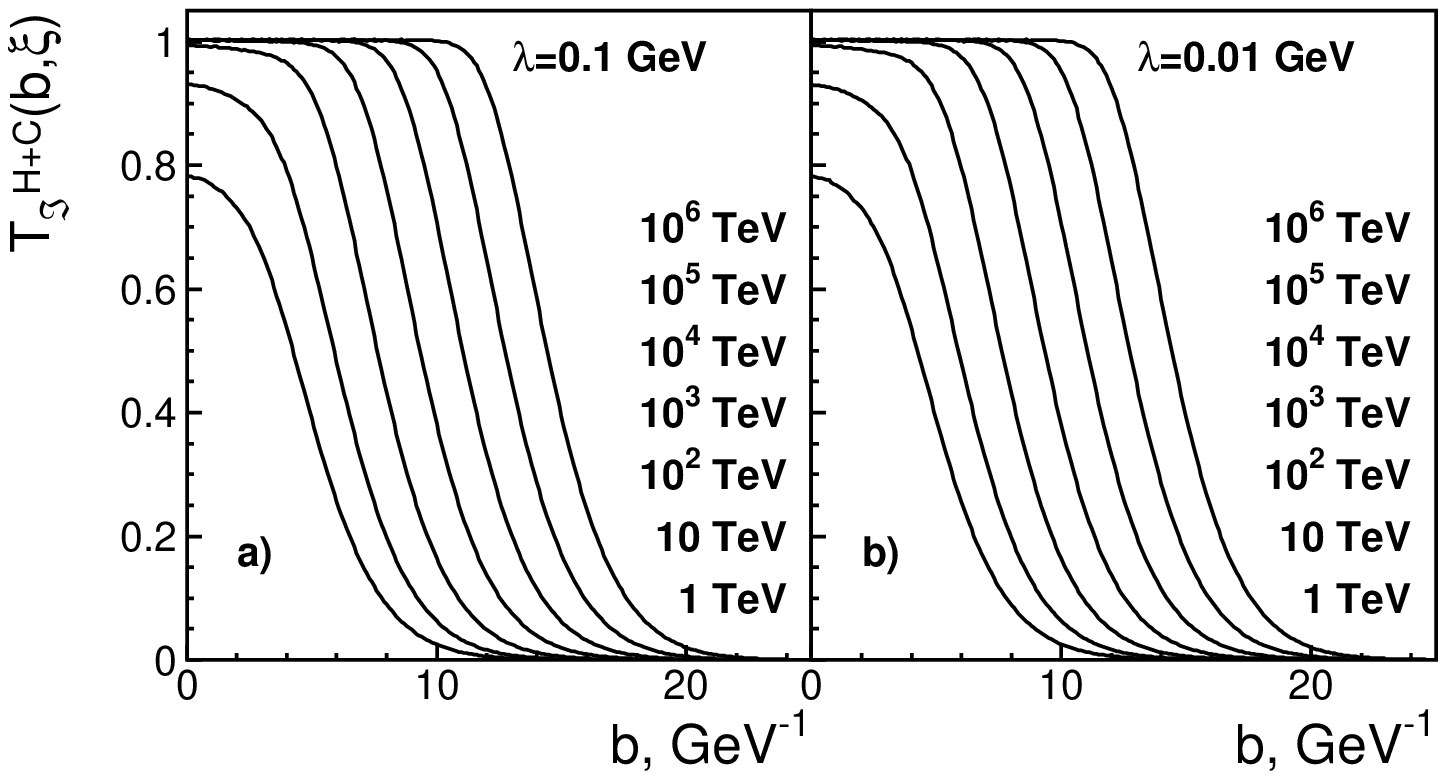,width=11cm} }
\centerline{
\epsfig{file=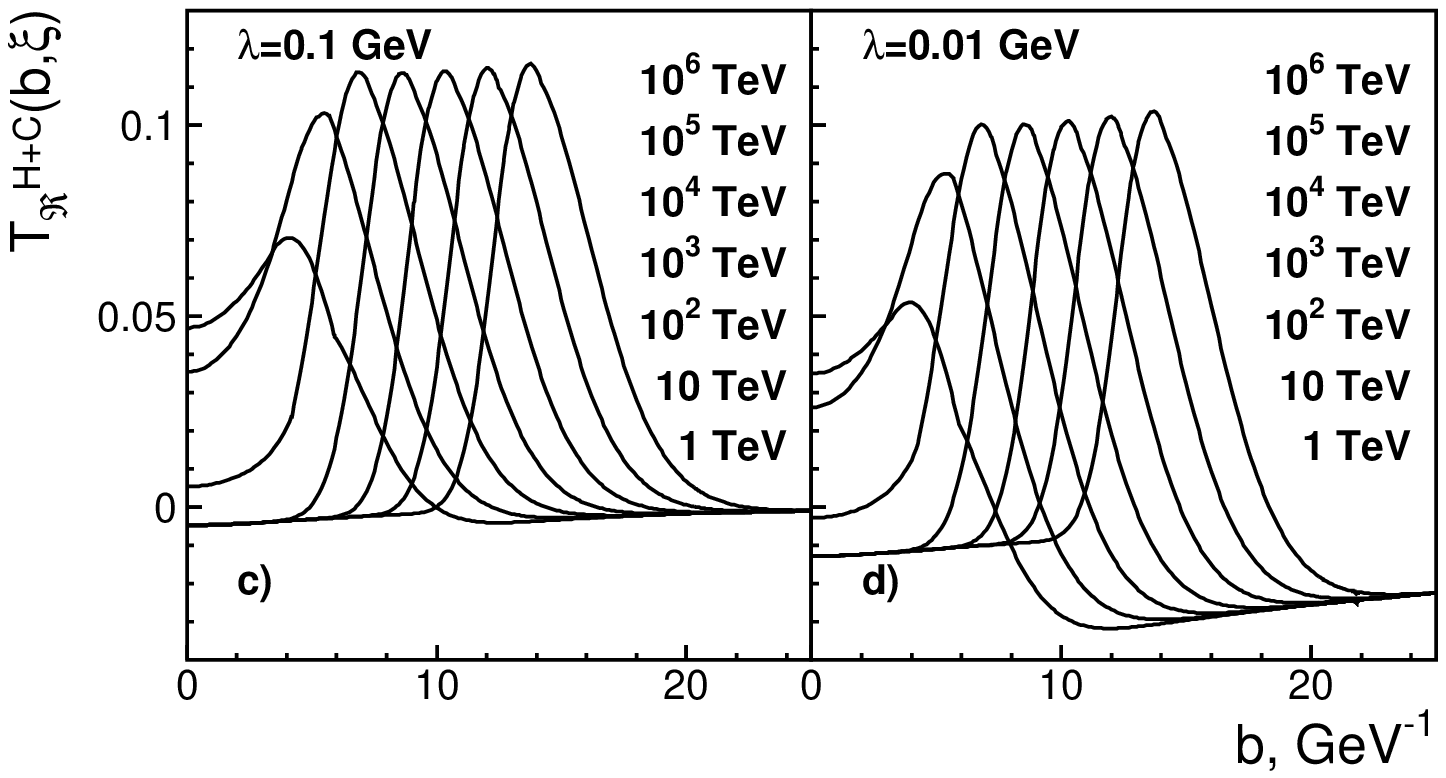,width=11cm}
}
\caption{
  Combined hadronic and coulombic interaction for the $pp$ scattering:
  Imaginary parts of the  profile functions at different $\lambda$
  at a set of energies $\sqrt{s}=1,10,10^2,...,10^6$ TeV; for
    $\sqrt{s}>100$ TeV the black disk mode is suggested.
}
\label{f6}
\end{figure*}

\begin{figure*}[ht]
%\Fig. 7
\centerline{\epsfig{file=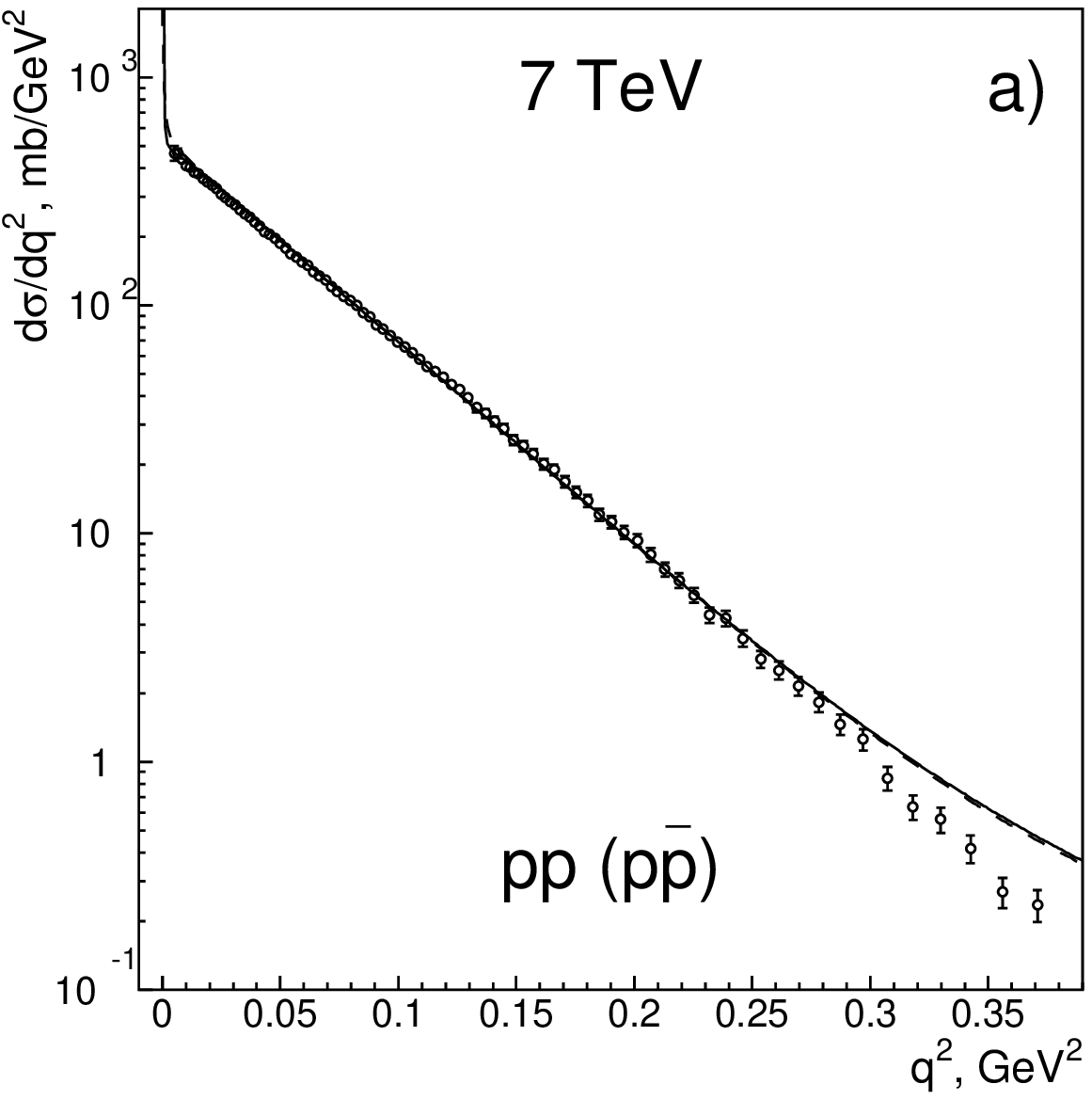,width=0.3\textwidth}
            \epsfig{file=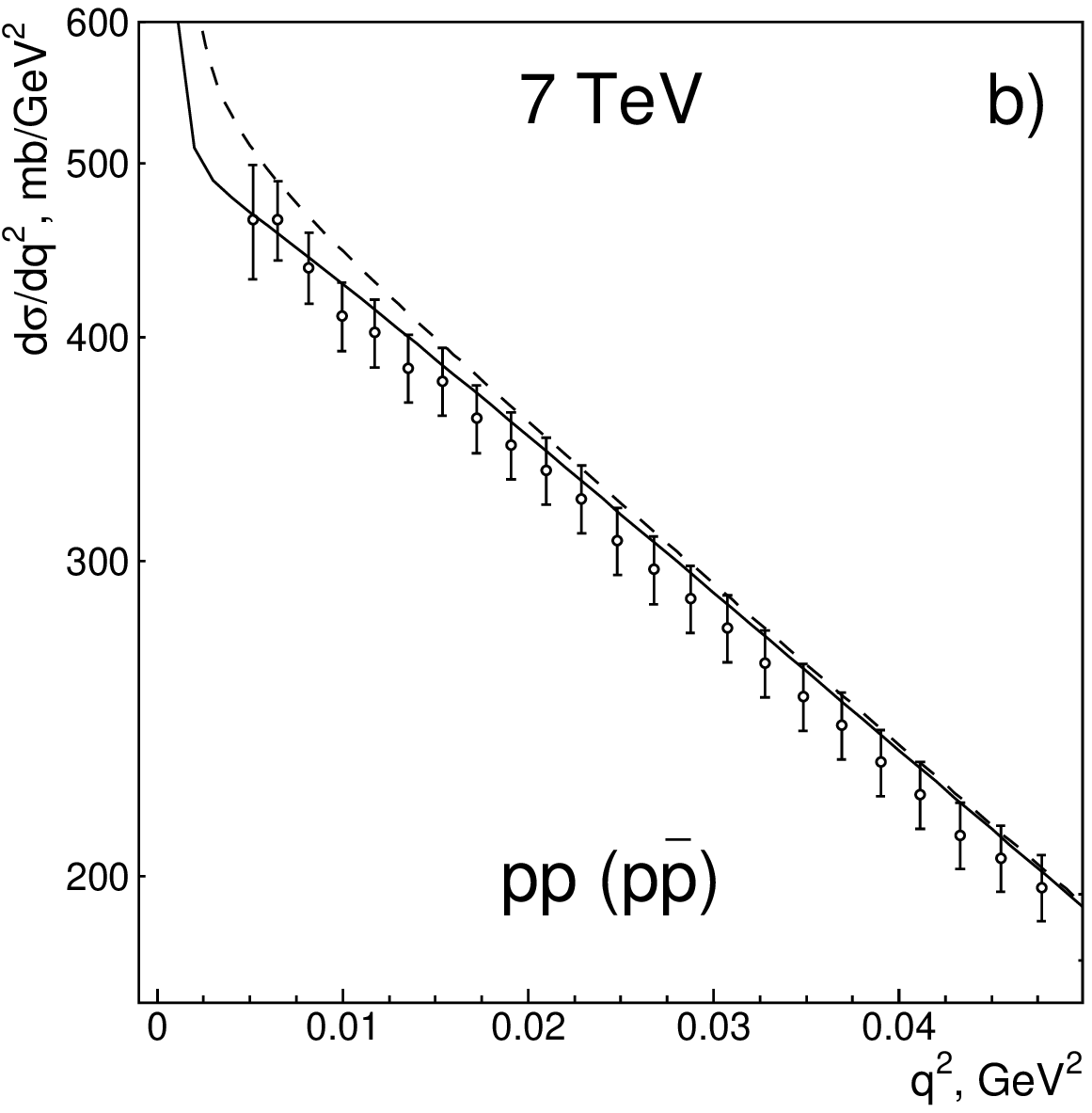,width=0.3\textwidth}}
\centerline{\epsfig{file=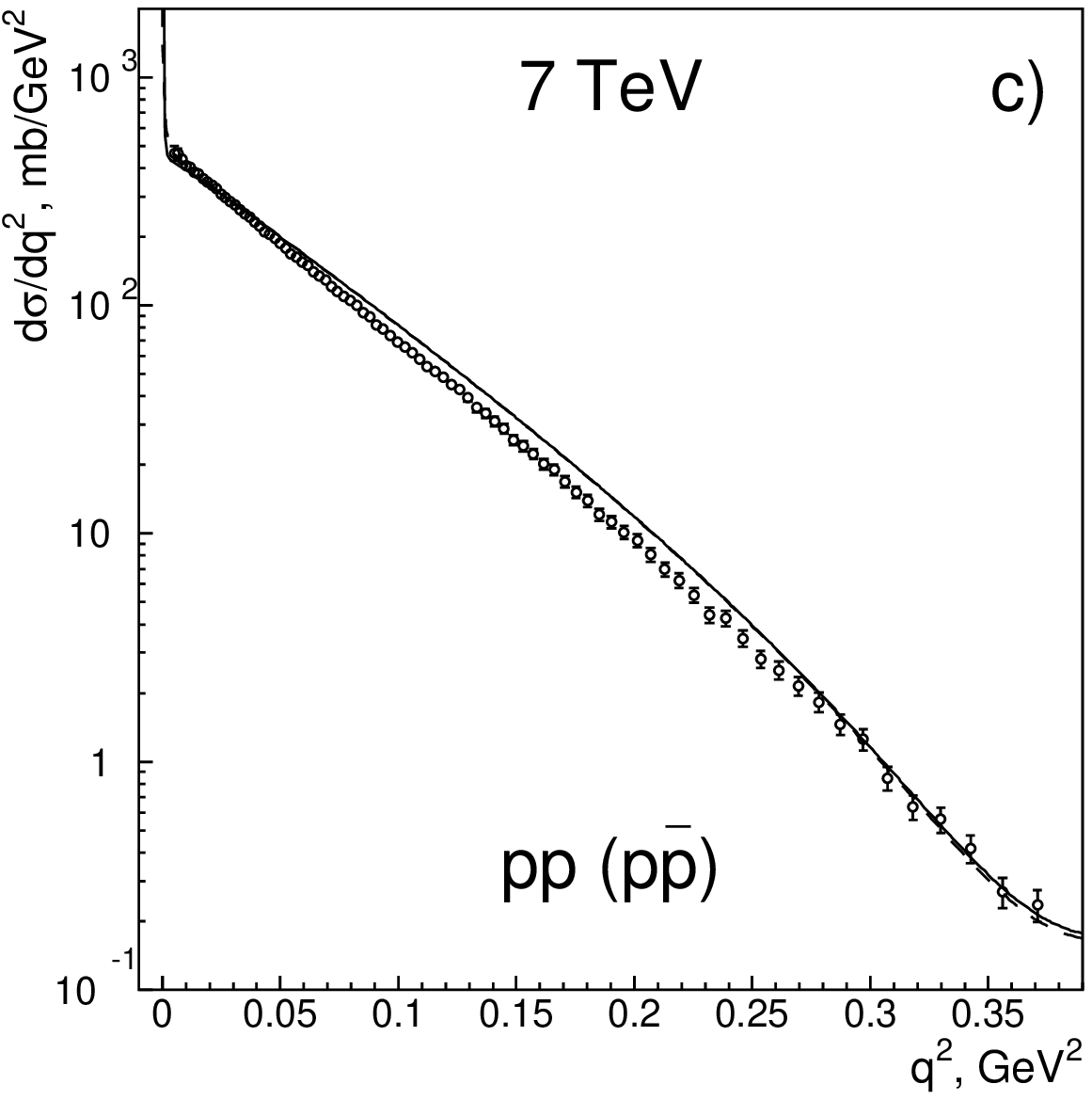,width=0.3\textwidth}
            \epsfig{file=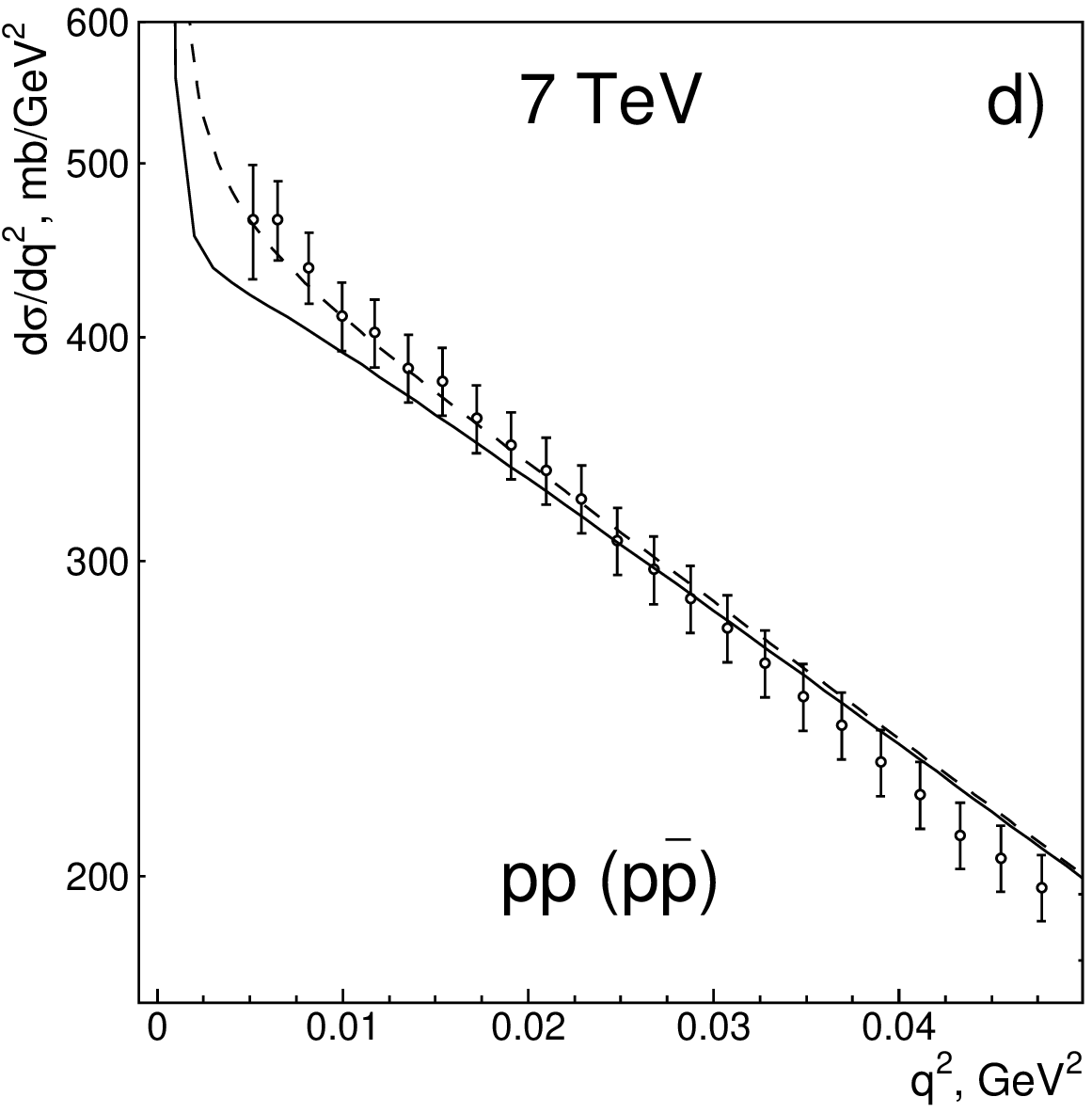,width=0.3\textwidth}}
\caption{
Diffractive scattering cross section for $pp$ at 7 TeV \cite{totem} versus
description with interplay of the Coulomb interaction (we use here
$\lambda$=0.01 GeV) and hadronic one (the real part of the hadronic
amplitude is taken into account): figures (a,b) refer to
version (1) for determination of the hadronic amplitude, figures (c,d) to
version (2); solid curves refer to $pp$, dashed ones to $p\bar p.$}
\label{fig-f2}
\end{figure*}

\subsection{Interplay of hadronic and Coulomb interactions in the $K$-matrix
function technique}

We consider two types of scattering amplitudes
and corresponding profile functions:
the amplitude with combined interaction taken into account,
$A^{C+H}({\bf q}^2 ,\xi)$ and $T^{C+H}(b,\xi)$,
and that with switched-off the Coulomb interaction,  following to (\ref{c1})
we use for them notation
$A({\bf q}^2 ,\xi)$ and $T(b,\xi)$. For the combined
interaction profile function we write:
\bea \label{6}
  T^{C+H}(b,\xi)&=&
\frac{-2iK^{C+H}(b,\xi)}{1-iK^{C+H}(b,\xi)}
\nn \\
&=&
\frac{-2i\left(K^{C}(b)+K(b,\xi)\right)}{1-i\left(K^{C}(b)
+K(b,\xi)\right)}\,,
\eea
and the Coulomb interaction is written as:
\bea
\label{13}
&&
A^C({\bf q}^2 )=
\pm i f_1({\bf q}^2)\frac{4\pi\alpha}{{\bf q}^2+\lambda^2}
f_2({\bf q}^2)\\
&&
-2i K^C(b)=
\pm i\int\frac{d^2q}{(2\pi)^2 }  e^{i{\bf q}{\bf b}}
f_1({\bf q}^2)
\frac{4\pi\alpha}{{\bf q}^2+\lambda^2} f_2({\bf q}^2)\,.
\nn
\eea
Here  $\alpha =1/137$; the upper/lower signs refer to the same/opposite
charges of the colliding particles.
The cutting parameter $\lambda$, which removes infrared
divergency, can be put to zero
in the final result for $A^{C+H}({\bf q}^2 ,\xi)$.
Colliding hadron form factors, $f_1({\bf q}^2)$ and $f_2({\bf q}^2)$,
guarantee the convergency of the integrals at ${\bf q}^2\to \infty$;
for the $pp^\pm$-collisions we use:
\be
f_1({\bf q}^2)=f_2({\bf q}^2)=\frac{1}{(1+\frac{{\bf q}^2}{0.71GeV^2})^2} \,.
\ee

\subsection{Numerical calculations}

In Fig. {\ref{f4}} we show $K^{C}(b) $
for $\lambda =0.1$ GeV and $0.01$ GeV.
The inclusion of the Coulomb interaction into consideration of hadron
diffractive scattering does not change the imaginary part of the
$K$-matrix function,
  $K_\Im^{H+C}(b,\xi)=K_\Im(b,\xi)$ . The real parts of the
  $K$-matrix functions $K_\Re^{H+C}(b,\xi)=K_\Re^{}(b,\xi)+K^{C}(b)$
for different $\lambda$ are shown in Fig. {\ref{f5}}.
% for $b<25$ GeV$^{-1}$. In the region $b>25$ GeV$^{-1}$ we have oscillatoric behavior
% inherent to Coulomb interaction, $K_\Re^{H+C}(b,\xi)\simeq K^{C}(b)$ (Fig.
% {\ref{f4}b}).

Imaginary and real parts of the profile functions, $T^{H+C}(b ,\xi)$
for the hadronic region, $b<25$ GeV$^{-1}$, are shown in Fig. {\ref{f6}};
considerable perturbations are seen in the real part.

With these $\lambda $'s we calculate at $\sqrt s=7$ TeV the profile function
$T^{C+H}(b,\xi_{LHC})$ and the corresponding amplitude $A^{C+H}({\bf q}^2 ,
\xi_{LHC})$. The determination of the hadronic amplitude, $A_\Im({\bf q}^2
,\xi_{LHC})$, is performed in terms of two versions:\\
1) with a direct application of the approximation (\ref{c5}) to the TOTEM
data \cite{totem},\\
2) using the the results of the Dakhno-Nikonov model \cite{DN,ann1}.

The description of the data for $\frac{d\sigma_{el}({\bf q}^2
,\xi_{LHC})}{d{\bf q}^2}$ in terms of these two
versions is shown in Fig. \ref{fig-f2}: here Figs. \ref{fig-f2}a,b
refer to the version (1) and
  Figs. \ref{fig-f2}c,d correspond to the version (2).
%Let us emphase the real part of the hadronic amplitude,
%given by eq. (\ref{e2}), is taken into account here.  Detailed presentation of different
%contributions into diffractive cross section can be found in Section 4.

The Dakhno-Nikonov model gives a somewhat worse description of the
$\frac{d\sigma_{el}({\bf q}^2 ,\xi_{LHC})}{d{\bf q}^2}$ at 7 TeV than that
using Eq. (\ref{c5}). This is not surprising because the model descibes
the data in a broad energy interval, 0.5-50 TeV \cite{ann2}, and the model
parameters are responsible for a complete set of the data.

% On Fig. {\ref{f4}a} we show $K^{C}(b) $ in the region of hadronic $b$; the region of
% very large $b$ is shown on Fig. {\ref{f4}b}. The form factor cutting of $A^C({\bf q}^2
% )$ results in freezing of the $K^{C}(b)$ at $\lambda< 0.001$ GeV (Fig. {\ref{f4}a}) while
% the infrared increase of the amplitude, $A^C({\bf q}^2 )\sim {\bf q}^{-2}$ at
% ${\bf q}^2\to 0$, turns into oscillatoric behavior of $K^{C}(b) $ at
% $\beta=\lambda b\ga 1$ (Fig. {\ref{f4}b}).
%
%
%The region
%of the coulombic impact parameters, $\beta=\lambda b\ga 1$, is shown
%for  $T^{H+C}(b ,\xi)$ on Fig. {\ref{f7}}. Oscillatoric structure of  $T^{H+C}(b ,\xi)$
%depends on energy weakly being determined mainly by  $K^{C}(b)$.

The specificity of the $K$-matrix function treating the amplitudes with
the Coulomb interaction is the use of the determination:
$K^C(b)= tg\delta^C(s,b)$.
Another determination was applied in \cite{kund}:
$K^C(b)=\delta^C(s,b)$.
%At present energies the difference
%is not essential within 10$\%$ accuracy however it can be more significant at
%$\sqrt{s}>10^2$ TeV.

\section{Conclusion}

On the basis of requirements of analyticity we calculate the leading terms
of the real part of the $pp$-scattering amplitude for diffractive
interactions at ultrahigh energies, $\sqrt{s}> 1$ TeV, and small momenta
transferred, ${\bf q}^2<0.4$ GeV$^2$.
We do not include into consideration the region with larger  ${\bf q}^2$:
in the region of larger momenta transferred mechanisms with conventional
Pomeron as well as short-range non-Pomeron interactions are possible (for
example, see Refs. \cite{koti,kang,levi,likh}) but here we concentrate our
attention on peripheral interactions.

In the region of the diffractive scattering cone the imaginary part of the
amplitude prevails over the real part
   $A_\Re ({\bf q}^2,\xi)/A_\Im ({\bf q}^2,\xi)\sim 1/\xi$.
The unitarity and analyticity requirements give unambiguously the leading
term of the real part.  We calculate $A_\Re ({\bf q}^2,\xi)$
at ${\bf q}^2\leq 0.4$ GeV$^2$ supposing for energies $\sqrt{s}> 100$ TeV
the black disk mode.

Presently we have a number of papers devoted to the asymptotic behavior of
diffractive amplitudes at ultrahigh energies, see for example
\cite{rMM,rAK,rOV} and references therein; in these papers, however, the
analyticity condition related to the $u$-channel is disregarded.

The interplay of the hadronic and Coulomb interactions at very small ${\bf
q_\perp}^2$ is discussed in terms of the $K$-matrix function. The
specificity of the scattering amplitude at ultrahigh energy is the dominance
of the mass-on-shall contributions in intermediate rescattering states
that results in the mass-on-shell origin of the $K$-matrix functions. Such
$K$-matrix functions allow to incorporate the Coulomb interaction terms
into the scattering amplitude straightforwardly, keeping te unitarity
condition. We present corresponding formulae and perform calculations for
the black disk mode.

\subsubsection*{Acknowledgement}

We thank Y.I. Azimov and A.V. Sarantsev for useful discussions.
  The work was supported by grants RFBR-13-02-00425 and
  RSGSS-4801.2012.2.

\end{document}